\documentclass[amsmath,amssymb,12pt,eps,amsfonts]{article}
\usepackage{epsfig,epic}
\usepackage{amsfonts,amssymb,amsmath,epsfig,epic,color}
\usepackage{graphicx}
\usepackage{dcolumn}
\usepackage{bm}
\newcommand{\beq}{\begin{equation}}
\newcommand{\eeq}{\end{equation}}
\newcommand{\beqs}{\begin{eqnarray}}
\newcommand{\eeqs}{\end{eqnarray}}

\begin{document}
\title{ \bf Non-relativistic Lee Model \\ in Three Dimensional Riemannian Manifolds }

\author{\centerline {\small  Fatih Erman$^1$, O. Teoman Turgut$^{1,\,2}$}
\\\and
 {\scriptsize{$^1$
Department of Physics, Bo\u{g}azi\c{c}i University, Bebek, 34342,
\.Istanbul, Turkey}}
\\\and
{\scriptsize{$^2$Feza G\"{u}rsey Institute, Kandilli, 81220
\.Istanbul, Turkey}}
\\
{\scriptsize{Electronic mail: turgutte@boun.edu.tr,
fatih.erman@boun.edu.tr }}}

\date{\scriptsize{\textsc{\today}}}

\maketitle

\abstract{In this work, we construct the non-relativistic Lee
model on some class of three dimensional Riemannian manifolds by
following a novel approach introduced by S. G. Rajeev
\cite{rajeevbound}. This approach together with the help of heat
kernel allows us to perform the renormalization non-perturbatively
and explicitly. For completeness, we show that the ground state
energy is bounded from below for different classes of manifolds,
using the upper bound estimates on the heat kernel. Finally, we
apply a kind of mean field approximation to the model for compact
and non-compact manifolds separately and discover that the ground
state energy grows linearly with the number of bosons $n$.}

\section{Introduction}

The Lee model, originally introduced in \cite{lee}, is an exactly
soluble and renormalizable model that includes the interaction
between a relativistic bosonic field (``pion") and a heavy source
(``nucleon") with one internal degree of freedom, which has two
eigenvalues distinguishing ``proton" and ``neutron". By heavy, we
mean that the recoil of the source is neglected. Although this
model is not very realistic, it reflects important features of
nucleon-pion system and presents  a powerful aspect that one can
do the renormalization without the use of perturbation techniques.
Moreover, the complete non-relativistic version of this model that
describes one heavy particle sitting at some fixed point
interacting  with a field of non-relativistic bosons is as
important as its relativistic counterpart. It is simpler than its
relativistic version because it is possible  to renormalize the
Hamiltonian of the system with only an additive renormalization of
the mass (energy) difference of the fermions. It has been studied
in a textbook by Henley and Thirring for small number of bosons
from the point of view of scattering matrix \cite{Thirring} and
there are further attempts in the literature from several point of
views \cite{nleepapers}. It is possible to look at the same
problem from the point of view of the resolvent of the Hamiltonian
in a Fock space formalism with arbitrary number of bosons (in fact
there is a conserved quantity which allows us to restrict the
problem to the direct sum of $n$ and $n+1$ boson sectors). This is
achieved in a very interesting unpublished paper by S. G. Rajeev
\cite{rajeevbound},
 in which a new non-perturbative formulation of renormalization has been
proposed. We are not going to review the ideas developed in there.
Instead, we suggest the reader to read through the paper
\cite{rajeevbound} to make the reading of this paper easier.

Following the original ideas developed in \cite{rajeevbound}, we
wish to extend the non-relativistic Lee model onto the Riemannian
manifolds with the help of heat kernel techniques, hoping that one
may understand the nature of renormalization on general curved
spaces better. In this work,  for the sake of simplicity we ignore
the motion of the heavy particle and take its position $a$ as a
given fixed point on the manifold. The construction of the model
is simply based on finding the resolvent of the regularized
Hamiltonian $H_\epsilon$ and show that a well-definite finite
limit of the resolvent exists as $\epsilon \rightarrow 0^+$
(called renormalization) with the help of heat kernel. We prove
that the ground state energy for a fixed number of bosons is
bounded from below, using the lower bound estimates of heat kernel
for some class of Riemannian manifolds, e.g., Cartan-Hadamard
manifolds (also explicitly $\mathbb{H}^3$), the minimal
submanifolds of $\mathbb{R}^3$ and closed compact manifolds with
nonnegative Ricci curvature. We also study the model in the mean
field approximation for compact and non-compact manifolds
separately and prove that the ground state energy grows linearly
with the number of bosons for both classes of manifolds.

The paper is organized as follows. In the first part, we construct
the model and show that the renormalization can be accomplished on
Riemannian manifolds. Then, we prove that there exists a lower
bound on the ground state energy. Finally, the model is examined
in the mean field approximation.

\section{The Construction of the Model}

We start with the regularized Hamiltonian of the non-relativistic
Lee model on a three dimensional Riemannian manifold
$(\mathcal{M},g)$ with a cut-off $\epsilon$. Adopting the natural
units ($\hbar= c=1$), one can write down the regularized
Hamiltonian on the local coordinates $x = (x_1,x_2,x_3)\in
\mathcal{M}$
\begin{equation}
H_\epsilon =H_0 + H_{I,\epsilon} \;,
\end{equation}
where
\begin{equation}
H_0=\int_{\mathcal{M}} d_g x \; \phi^\dag(x)\left(-{1\over
2m}\nabla^2_g+m \right)\phi(x) \;,
\end{equation}
\begin{equation}
H_{I,\epsilon}=\mu(\epsilon){1-\sigma_3\over
2}+\lambda\int_{\mathcal{M}} d_g x \;
\rho_\epsilon(x,a)\left(\phi(x)\, \sigma_-+\phi^\dag(x) \,
\sigma_+ \right) \;. \end{equation}
Here, $d_g x = \sqrt{\det g_{i j}}\, d x $ is the volume element
and $\nabla^2_g$ is Laplace-Beltrami operator or simply Laplacian,
and $\phi^{\dag}(x)$, $\phi(x)$ is the bosonic
creation-annihilation operators defined on the manifold with the
metric structure $g_{ij}$. Sometimes we shall write $\phi_g(x)$ in
order to specify which metric structure it is associated with but
for now we simply write down $\phi(x)$. Also, $
\rho_\epsilon(x,a)$ is a family of functions which converge to the
Dirac delta function $\delta_g (x,a)$ (with the normalization
$\int_{\mathcal{M}} d_g x \; \delta_g (x) =1 $) around the point
$a$ on $\mathcal{M}$ as we take the limit $\epsilon \to 0^+$. The
Pauli spin matrices $\sigma_{\pm} = {1\over 2} (\sigma_1 \pm i
\sigma_2 )$ and $\sigma_3$ are regarded as a matrix representation
of the fermionic creation and annihilation operators acting on
$\mathbb{C}^2$ and $\mu(\epsilon)$ is a bare mass difference
between the ``proton" and ``neutron" states of the two state
system (``nucleon"). Its explicit form will be determined later
on. Although the number of bosons is not conserved in the model,
one can derive from the equations of motion that there exists a
conserved quantity
\begin{equation} \nonumber
Q = - {1- \sigma_3 \over 2} + \int_{\mathcal{M}} d_g x \;
\phi^{\dag}(x) \phi(x)
\end{equation}
which takes only positive integer values. If $Q=n \in
\mathbb{Z}^+$, we have spin-up state (``proton") with $n$ bosons
or spin-down state (``neutron") with $n-1$ bosons. We can think of
the latter as a bound state of the system. If one considers the
Hamiltonian without the cut-off $\epsilon$, it can be shown that
the bound state energy diverges. Before discussing how to deal
with the infinities, we must see how the infinities emerge in our
model. The simplest way to realize this is just to look at the
sector that contains the ``neutron" or the ``proton" with one
boson, which corresponds to $Q=1$, that is, we propose the
following eigenstate ansatz \cite{rajeevlecture}:
\begin{equation} \label{ansatz}
| u,  \psi \rangle = \left(%
\begin{array}{c}
   \int_{\mathcal{M}} d_g x \; \psi(x) \phi^{\dag} (x) |0 \rangle   \\
 u |0 \rangle \\
\end{array}%
\right)\;.
\end{equation}
For simplicity we explicitly perform our calculations for compact
manifolds here, but our result is also valid for non-compact
manifolds, which we are interested in. If the manifold
$\mathcal{M}$ is compact, then the Laplacian has a discrete
spectrum and there is a family of orthonormal complete
eigenfunctions $f_\sigma(x)\in L^2 (\mathcal{M})$ \cite{Rosenberg}
satisfying,
\begin{eqnarray} -\nabla_g^2 f_\sigma(x)&=& \sigma f_\sigma(x) \;,\cr
\int_{\mathcal{M}} d^3_g x \; f_{\sigma}^{*}(x)\,
f_{\sigma'}(x)&=& \delta_{\sigma \sigma'}  \;, \cr \sum_{\sigma}
f_{\sigma}^{*}(x) f_\sigma(y)&=& \delta_g(x,y) \;.
\end{eqnarray}
In fact, one can extend these expressions to some noncompact
manifolds, and they also satisfy these properties by an
appropriate generalization of the measures to the continuous
distributions in the sense of \cite{berezansky}.

From the eigenvalue equation $H |u, \psi \rangle = E |u, \psi
\rangle$, we find the set of equations in terms of the bosonic
wave function $\psi(x)= \sum_\sigma f_\sigma (x) \psi(\sigma)$
\begin{eqnarray} \label{psi}
\psi (\sigma)&=& {u \lambda f^{*}_{\sigma}(a) \over E - \left(
{\sigma \over 2 m} + m \right)}\;, \\ \label{diveq} u(E-\mu)&=&
\lambda \sum_{\sigma} f_{\sigma}(a) \psi(\sigma)\;.
\end{eqnarray}
If we substitute the equation (\ref{psi}) into the equation
(\ref{diveq}), we obtain
\begin{eqnarray} \nonumber
\mu = E - \lambda^2 \sum_\sigma {|f_{\sigma} (a)|^2 \over E -
\left( {\sigma \over 2 m} + m \right)}\;.
\end{eqnarray}
Expressing this equation in terms of heat kernel $K_{s/2m}(x,x')=
\langle x | e^{s \nabla_{g}^{2}/2m} |x' \rangle$, we get
\begin{eqnarray} \label{divint}
\mu = E + \lambda^2 \int_{0}^{\infty} d s \; K_{s/2m} (a,a)\,
e^{-s(m-E)}\;.
\end{eqnarray}
Let us make a short  digression  on heat kernels. When we want to
emphasize the metric structure $g_{ij}$ on which the heat kernel
is based, we shall use the notation $K_{s/2m}(x,x';g)$ throughout
the paper. Although the notion of heat kernel can be defined on
any Riemannian manifold, the explicit formulas only exist for some
special class of manifolds, for example, Euclidean spaces
$\mathbb{R}^{d}$ \cite{grigoryan} and hyperbolic spaces
$\mathbb{H}^d$ \cite{heatkernelh3}. Some of the well known
properties of the heat kernel on $\mathcal{M}$ in dimensionless
parameters $t$ and $x$ are
\begin{eqnarray}
K_t (x,y) &=& K_t (y,x) \hspace{2cm} \mathrm{Symmetry \;
property}\;, \cr {\partial K_t(x,x')\over
\partial t}-\nabla^2_g K_t(x,x')&=& 0 \hspace{3cm} \mathrm{Heat \;
equation}\;, \cr  \lim_{t\to 0^+} K_t(x,x')&=&\delta_g(x,x')
\hspace{2cm} \mathrm{Initial \; condition}\;,\cr
\int_{\mathcal{M}} d_g x \; K_{t_1}(x,z)K_{t_2}(z,y)&=& K_{t_1 +
t_2}(x,y), \hspace{1cm} \mathrm{Reproducing \; property} \;,
\end{eqnarray}
for $t \geq 0$ only. If in addition $\mathcal{M}$ is a compact
manifold, we have
\begin{equation}
K_t(x,y)= \sum_{\sigma} e^{-t \sigma} f_{\sigma}^{*}(x)
f_\sigma(y)\; ,
\end{equation}
which converges uniformly on $\mathcal{M} \times \mathcal{M}$ .
Again we assume that this has a proper analogous expression for
non-compact manifolds we are interested in. When the manifold
$\mathcal{M}$ is a complete Riemannian manifold with Ricci
curvature bounded from below then the heat kernel satisfies the
stochastic completeness property \cite{grigoryan, yau}:
\begin{equation} \nonumber
\int_{\mathcal{M}} d_g x \; K_t(x,y)=1
\end{equation}
On a compact manifold stochastic completeness is always satisfied
\cite{chavel}.

The integral in the equation (\ref{divint}) diverges due to the
asymptotic expansion of the diagonal part of heat kernel near
$s=0$ for any three dimensional geodesically complete manifold
whose injectivity radius has a positive lower bound
\cite{schoenyau}
\begin{equation} \label{asymheat}
\lim_{s\to 0^+} K_{s/2m}(a,a) \sim {1\over (4\pi s/2m)^{3/2}}
\sum_{k=0}^{\infty} u_k(a,a) (s/2m)^k \;.
\end{equation}
Here the $u_k(a,a)$ are functions given in terms of curvature
tensor of the manifold and its covariant derivatives at the point
$a$. As a result of this, the bound state energy becomes
divergent. In flat spaces, one can do the similar calculations in
momentum space and find that \cite{rajeevlecture}
\begin{equation} \label{r3momentum}
\mu = E - \lambda^2 \int { d p \over E - \omega(p)}
\end{equation}
where $\omega(p)= {p^2 \over 2m} + m $. This momentum integral
blows up at large values of momentum in three dimensions. The
problem is basically taking the integral over all momenta because
our no-recoil approximation breaks down for large enough momenta.
So, we introduce an ultraviolet cut-off $\Lambda$ in the upper
bound of the integral. Since large momenta means small distances,
this cut-off corresponds to putting a small distance cut-off in
coordinate space. Performing the calculations in coordinate space
one see that small distance cut-off can be replaced with a short
``time" cut-off $\epsilon$ in the lower limit of the integral
(\ref{divint}). Here we  show that the idea of short ``time"
cut-off  will work  on Riemannian manifolds, whereas the momentum
cut-off is not a natural method to use.

Therefore, we first regularize the Hamiltonian, that is, introduce
the cut-off $\epsilon$ on the lower bound of the integral and make
the parameters in the Hamiltonian depend on $\epsilon$ such that
all physical quantities are independent of it. So we define
\begin{eqnarray} \label{mueps}
\mu(\epsilon) = \mu + \lambda^2 \int_{\epsilon}^{\infty} d s \;
K_{s/2m} (a,a)\, e^{-s(m-\mu)}\;,
\end{eqnarray}
where $E$ is traded with  $\mu$ which is defined as the physical
energy of the composite state which consists of a boson and the
attractive heavy neutron at the center. Now, using (\ref{psi}),
(\ref{diveq}) and (\ref{mueps}), we get the finite expression for
the bound state energy
\begin{equation}
E = \mu + \lambda^2 \int_{0}^{\infty} d s \; K_{s/2m} (a,a)
\left[e^{-s(m-\mu)} - e^{-s(m-E)}\right] \;.
\end{equation}
Here it is easy to see that $E = \mu$ is a possible solution to
this equation. Then, one can calculate the bosonic wave function
$\psi(x)$ for $E=\mu$
\begin{equation}
\psi(x)= - u \lambda \int_{0}^{\infty} d s \; K_{s/2m} (x,a)
e^{-s(m-\mu)}\;, \label{psist}
\end{equation}
if $x\neq a$. Although the wave function is divergent as $x
\rightarrow a$, it is square integrable as we will see. If we
substitute (\ref{psist}) into our ansatz (\ref{ansatz}), we get
\begin{equation}
|u \psi \rangle = u \left(%
\begin{array}{c}
   -{\lambda \over H_0 - \mu} \phi^{\dag} (a) |0 \rangle   \\
  |0 \rangle \\
\end{array}%
\right) \;. \label{eigenvector1}
\end{equation}
Normalizability of (\ref{eigenvector1}) can be easily seen by
using the properties of heat kernel \footnote{The same
normalization can also be found by writing the operators $\phi(a)$
in the eigenbasis  $ f_\sigma (a)$ of the Laplacian.}:
\begin{eqnarray} \label{norm} &\ &
\lambda^2 \int_{\mathcal{M}} d_g x \int_{0}^{\infty} \; d s_1 \; d
s_2 K_{s_{1}/2m}(x,a) K_{s_{2}/2m}(x,a)  \; e^{-(s_1 + s_2)(m -
\mu)} \cr &\ & \hskip-2cm =  \int_{0}^{\infty} \; d s  \left({1
\over 2}\int_{-s}^{s} d t \right) K_{s/2m}(a,a) \; e^{-s(m - \mu)}
= \int_{0}^{\infty} \; s \,d s \;   e^{-s(m - \mu)} \,
K_{s/2m}(a,a) \; \;,
\end{eqnarray}
and as a result we find the normalization to be
\begin{equation} \label{norm}
\left[ 1+ \lambda^2 \int_{0}^{\infty} s \, d s \; e^{-s(m - \mu)}
K_{s/2m}(a,a) \right]^{-1/2}\;.
\end{equation}
This integral is finite due to the short and long time behaviour
of heat kernel.

One can also consider the scattering of a boson from the ``proton"
at rest {\it{on a noncompact manifold}} \footnote{Compact
manifolds have only discrete spectrum.}. The inhomogeneous
Schr\"{o}dinger equation $(H -E)| u, \psi \rangle = |v,\chi
\rangle$ leads to
\begin{equation}
\psi (\sigma) = { \chi(\sigma) - \lambda u f^{*}_{\sigma} (a)\over
{\sigma \over 2 m} + m -E } \label{psichi}
\end{equation}
and
\begin{equation} \label{scatt}
\lambda \sum_\sigma f_\sigma (a) \psi(\sigma) + u(\mu-E) = v \;,
\end{equation}
If we substitute (\ref{psichi}) into (\ref{scatt}), we find
\begin{equation} \nonumber
-\lambda^2 u \sum_\sigma {|f_\sigma (a)|^2 \over {\sigma \over 2
m} + m -E} + u(\mu-E) = v - \lambda \sum_\sigma {f_\sigma (a)
\chi(\sigma) \over {\sigma \over 2 m} + m -E}\;.
\end{equation}
If we express the above equation in terms of heat kernel, we
immediately see that the integral is divergent due to singularity
near $s=0$. So we must do the regularization by introducing a
cut-off $\epsilon$ in the lower limit of the integral and choose
$\mu(\epsilon)$ as above. Taking the limit $\epsilon\rightarrow
0^+$ we get the following finite expression
\begin{eqnarray} \nonumber
&\ & \lim_{\epsilon\rightarrow 0^+ } \left( \lambda^2 u
\int_{\epsilon}^{\infty} d s \; K_{s/2m} (a,a)\, e^{-s(m-\mu)} -
\lambda^2 u \int_{\epsilon}^{\infty} d s \; K_{s/2m} (a,a) \,
e^{-s(m-E)} \right) \cr  &\ & \hspace{2cm} = v - \lambda
\int_{\mathcal{M}} d_g x \;\chi(x) \left(\int_{0}^{\infty} d s \;
K_{s/2m}(x,a)\, e^{-s(m-E)} \right) \;.
\end{eqnarray}
From the above equation, it follows that
\begin{eqnarray} \nonumber
&\ & u \equiv u[v,\chi]= \left[ \lambda^2 \int_{0}^{\infty} d s \;
K_{s/2m} (a,a)\, \left(e^{-s(m-\mu)} - e^{-s(m-E)}\right)
\right]^{-1} \cr &\ & \hspace{3cm} \times \left[v - \lambda
 \int_{0}^{\infty} d s \;
\left( \int_{\mathcal{M}} d_g x \;K_{s/2m}(x,a)\chi(x)\right) \,
e^{-s(m-E)}\right] \;.
\end{eqnarray}
We can also read off  $\psi(x)$ from the equation (\ref{psichi})
when $x\neq a$
\begin{equation} \label{scatpsi}
\hskip-0.2cm \psi(x)= \int_0^\infty d s \; e^{-s(m-E)}\int d_g y
\, K_{s/2m}(x,y) \, \chi(y)-\lambda \, u[v,\chi] \int_0^\infty d s
\; K_{s/2m}(x,a)\, e^{-s(m-E)} \;,
\end{equation}
and $\psi(a)=\lambda^{-1}( v-u[v,\chi](\mu-E))$ as a result of the
equation (\ref{scatt}). It is important to remind the reader that
these expressions should be understood in the sense of analytic
continuation in the complex $E$-plane to their largest domain of
definition. Indeed if the real part of $E$ is smaller than $m$
these integrals all make sense, and the resulting expressions are
just the Green's function, or the resolvents for the Laplace
operator, which exists away from  the positive real axis.

Up to now, we have shown that non-relativistic Lee model on a
manifold is divergent and can be renormalized with the help of
heat kernel by considering the problem for $Q=1$ sector. However,
it  is not clear in this formulation, how we can extend this
method to the case that have arbitrary number of particles, say
$Q=n$  sectors and show that the ground  state energy is bounded
from below. Nevertheless, we have an alternative and powerful
method which is developed by Rajeev \cite{rajeevbound}. From now
on, we will follow his approach in order to construct and develop
the model on a manifold for any sector.

Let us first express the regularized Hamiltonian as a $2\times 2$
block split according to $\mathbb{C}^2$:
\begin{equation} H_\epsilon -E =
\left(%
\begin{array}{cc}
  H_0-E & \lambda\int_{\mathcal{M}} d_g x \; \rho_\epsilon(x,a)\, \phi^\dag(x) \\
   \lambda\int_{\mathcal{M}} d_g x \; \rho_\epsilon(x,a)\, \phi(x) & H_0- E
   +\mu(\epsilon)
\end{array}%
\right)\;.
   \end{equation}
Then, one can construct the regularized resolvent of this
hamiltonian using an explicit formula given by Rajeev
\cite{rajeevbound}
\begin{equation}
R_\epsilon(E)= {1 \over H_\epsilon -E}= \left(%
\begin{array}{cc}
  \alpha & \beta^\dagger \\
  \beta & \delta \\
\end{array}%
\right)\;,
\end{equation}
where
\begin{eqnarray}
\alpha&=&{1\over H_0-E}+{1\over H_0-E} \; b^\dag \;
\Phi_\epsilon^{-1} (E) \; b \; {1\over H_0-E}\cr \beta&=&-
\Phi^{-1}_\epsilon(E)\; b \; {1\over H_0-E}\cr
\delta&=&\Phi^{-1}_\epsilon(E)\cr b&=&\lambda \int_{\mathcal{M}}
d_g x \; \rho_\epsilon(x,a)\,\phi(x) \;.
\end{eqnarray}
Here \textit{$E$ should be considered as a complex variable}. Most
importantly, the operator $\Phi_\epsilon(E)$, called principal
operator, is given as
\begin{equation}
\Phi_\epsilon(E)=H_0-E+\mu(\epsilon)-\lambda^2\int_{\mathcal{M}}
d_g x \, d_g y \; \rho_\epsilon(x,a) \rho_\epsilon(y,a)\;
\phi(x){1\over H_0-E}\phi^\dagger(y) \;,\label{phi_e}
\end{equation}
Once we have a proper definition of the principal operator, all
the divergences are removed since the resolvent is expressed in
terms of it. We can extend the principal operator by analytic
continuation to its largest domain of definition in  the complex
energy plane. For our purposes, we will assume that $\Re(E) < n m
+ \mu$. In fact, the energy of bound states are interesting and
they satisfy the required conditions as we will see. Now, we will
do the normal ordering of the operators in (\ref{phi_e}) by using
the commutation relations of the operators $\phi(x)$ and
$\phi^{\dag}(x)$
\begin{equation} \nonumber
\phi(x){1\over H_0-E}=\int_{\mathcal{M}} d_g x \int_0^\infty d s
\; e^{-s(H_0-E)}\, K_{s/2m}(x,x') \, \phi(x') \;,
\end{equation}
which can be proved simply by using an eigenfunction expansion.
Then, the principal operator can be written in terms of heat
kernel
\begin{eqnarray} \nonumber
\Phi_\epsilon(E)&=& H_0-E-\lambda^2\int_0^\infty d s
\int_{\mathcal{M}} d_g x\, d_g y\, d_g x'\, d_g y' \;
\rho_\epsilon(x,a)\rho_\epsilon(y,a)
 \cr & \ & \hspace{1cm} \times \; K_{s/2m}(x,x') \, K_{s/2m}(y,y') \, \phi^\dag(y')\,
e^{-s(H_0+2m-E)} \, \phi(x')\cr &\ & \hskip-2cm  + \;
\mu(\epsilon)-\lambda^2\int_0^\infty d s \int_{\mathcal{M}} d_g y
\, d_g y' \; K_{s/2m}(y,y')\rho_\epsilon(y,a)\rho_\epsilon(y',a)\,
e^{-s(H_0+m-E)} \,.
\end{eqnarray}
Since the heat kernel is a natural delta sequence, we can set
$\rho_\epsilon(x,a) =  K_{\epsilon/4m}(x,a)$ without loss of
generality. Hence,
\begin{eqnarray}\nonumber
\Phi_\epsilon(E)& = & H_0 - E - \lambda^2 \int_0^\infty d s
\int_{\mathcal{M}} d_g x \, d_g y \, d_g x' \, d_g y' \;
K_{\epsilon/4m}(x,a)K_{\epsilon/4m}(y,a)  \cr &\ & \hspace{1cm}
\times \; K_{s/2m}(x,x') \,K_{s/2m}(y,y')\, \phi^\dag(y') \,
e^{-s(H_0+2m-E)}\, \phi(x') \cr & \ & \hskip-2.5cm
 + \; \mu(\epsilon)-\lambda^2 \int_0^\infty d s \int_{\mathcal{M}} d_g y \, d_g y'
  \;
K_{s/2m}(y,y')K_{\epsilon/4m}(y,a)K_{\epsilon/4m}(y',a)\,
e^{-s(H_0+m-E)}\;. \label{phi2}
\end{eqnarray}
Using the reproducing property of the heat kernel, one can rewrite
the above equation as
\begin{eqnarray} \nonumber
\Phi_\epsilon(E)&=&H_0-E-\lambda^2\int_0^\infty d s
\int_{\mathcal{M}} d_g x \, d_g y \;
K_{(2s+\epsilon)/4m}(x,a)K_{(2s+\epsilon)/4m}(y,a) \cr &\ &
\hskip-2cm \times \; \phi^\dag(x)\, e^{-s(H_0+2m-E)}\, \phi(y) +
 \mu(\epsilon)-\lambda^2\int_0^\infty d s \;
 K_{(s+\epsilon)/2m}(a,a)\,
e^{-s(H_0+m-E)} \;.
\end{eqnarray}
Shifting the variable $s$ in the first integral by $\epsilon/2$
and in the second integral by $\epsilon$, we get
\begin{eqnarray} \nonumber
\Phi_\epsilon(E)&=&H_0-E-\lambda^2\int_{\epsilon/2}^\infty d s
\int_{\mathcal{M}} d_g x \, d_g y \; K_{s/2m}(x,a)K_{s/2m}(y,a)
\cr &\ & \hskip-2cm \times \; \phi^\dag(x)\,
e^{-(s-\epsilon/2)(H_0+2m-E)}\, \phi(y) +
\mu(\epsilon)-\lambda^2\int_\epsilon^\infty d s \; K_{s/2m}(a,a)\,
e^{-(s-\epsilon)(H_0+m-E)} \;.
\end{eqnarray}
If we take the limit $\epsilon\to 0^+$, the only divergence is
coming from the lower limit of the second integral term due to
singular behavior of the diagonal part ($k=0$ term in the sum) of
heat kernel near $s=0$, see equation (\ref{asymheat}).

One can also see that the first interaction term is actually
finite due to the quite sharp bounds on the heat kernel for
various classes of manifolds \cite{grigoryan,liyau}. In fact we
will explicitly show later on that this term is really a finite
expression by working out a bound on the  spectrum of the model.
Since the principal operator or resolvent is not well defined in
this limit, we must now regularize the model by choosing
$\mu(\epsilon)$ exactly the same as in (\ref{mueps}). Then, we
find
\begin{eqnarray} \nonumber
&\ & \Phi_\epsilon(E)= H_0-E-\lambda^2 \int_{\epsilon/2}^\infty d
s \int_{\mathcal{M}} d_g x \, d_g y \; K_{s/2m}(x,a)K_{s/2m}(y,a)
\cr &\ & \hskip-1.5cm \times \; \phi^\dag(x)
e^{-s(H_0+2m-E)}\phi(y) +
 \mu+\lambda^2  \int_\epsilon^\infty d s \;
K_{s/2m}(a,a)\left[e^{-s(m-\mu)}-e^{-s(H_0+m-E)}\right]\;.
\end{eqnarray}
Here the limit $\epsilon\to 0^+$ is now well-defined so we have
\begin{eqnarray}
\Phi(E)&=&H_0-E+\mu+\lambda^2 \int_0^\infty d s \;
K_{s/2m}(a,a)\left[e^{-s(m-\mu)} -e^{-s(H_0+m-E)}\right]\cr & \ &
\hskip-1cm -\lambda^2 \int_0^\infty d s \int_{\mathcal{M}} d_g x
\, d_g y \; K_{s/2m}(x,a)K_{s/2m}(y,a) \, \phi^\dag(x)
e^{-s(H_0+2m-E)}\phi(y) \;. \end{eqnarray}
This is the renormalized form of the principal operator so that we
have a well-defined explicit formula for the resolvent of the
Hamiltonian in terms of the inverse of the principal operator
$\Phi^{-1}(E)$.

The spectrum of the Hamiltonian is the set of numbers $E$ at which
the resolvent does not exist (discrete spectrum) or exist but is
unbounded (continuous spectrum). Thus, the continuous spectrum is
that of $H_0$ and the values of $E$ where  $\Phi(E)$ does not have
a bounded inverse. Since, there are no
poles in ${1 \over H_0 - E }$, the poles corresponding to bound
states must arise from those of $\Phi^{-1}(E)$, that is, the roots
of the equation 
\begin{equation} \label{spectrum}
\Phi(E) |\Psi \rangle = 0 \;,
\end{equation}
determine the poles in the resolvent, which means that the principal operator
$\Phi(E)$ determines the bound state spectrum of the theory. 
After we have found a root, we can determine the corresponding
eigenstate of the Hamiltonian. A trivial example is the bosonic
vacuum state,
\begin{eqnarray} \nonumber
\Phi(E)|0 \rangle&=& \bigg\{ H_0-E+\mu+\lambda^2 \int_0^\infty d s
\; K_{s/2m}(a,a)\left[e^{-s(m-\mu)} -e^{-s(H_0+m-E)}\right]\cr & \
& \hskip-2.5cm -\lambda^2 \int_0^\infty d s \int_{\mathcal{M}} d_g
x \, d_g y \; K_{s/2m}(x,a)K_{s/2m}(y,a) \, \phi^\dag(x)
e^{-s(H_0+2m-E)}\phi(y)\bigg\} |0 \rangle = 0 \;,
\end{eqnarray}
where the root can be easily found to be
\begin{equation} \nonumber
E=\mu \,.
\end{equation}
We remark here that the linear eigenvalue
 problem is converted into a nonlinear problem for
an operator family parametrized  through the energy eigenvalues
after the renormalization.

The residue of the pole in the resolvent is the projection
operator to the corresponding eigenspace of $H$ \cite{simon}
\begin{equation}
\mathbb{P}_{\mu}= {1 \over 2 \pi i} \oint_{\Gamma_\mu} d E \; R(E)
\;,
\end{equation}
where $\Gamma_\mu$ is a small contour enclosing the isolated
eigenvalue $\mu$ in the complex plane. For this contour integral
we need to evaluate the residue
\begin{equation}
\mathrm{Res}_{E= \mu} (\Phi^{-1}(E)|0 \rangle \langle 0 |) =
(\Phi'(\mu))^{-1} |0 \rangle \langle 0 | \;.
\end{equation}
As a result of this calculation, we find the projection operator
to be
\begin{eqnarray}
\mathbb{P}_{\mu} &=& \left[1 + \lambda^2 \int_{0}^{\infty} s \, d
s \; e^{-s(m-\mu)} K_{s/2m}(a,a)\right]^{-1} \cr &\ & 
\hspace{1cm}\times \left(
\begin{array}{cc}
  {\lambda  \over H_0 - \mu} \phi^{\dag}(a)|0 \rangle \langle 0 | \phi(a)  {\lambda  \over H_0 - \mu} &  - {\lambda \over H_0 - \mu}
  \phi^{\dag}(a)|0 \rangle \langle 0 | \\
 -|0 \rangle \langle 0 |\phi(a) {\lambda  \over H_0 - \mu} & |0 \rangle \langle 0 | \\
\end{array}%
\right)\;.
\end{eqnarray}
Thus, one can read off the eigenvector of $H$ corresponding to the
root $E= \mu$ from the projection operator and then find the
normalizable eigenstate (\ref{eigenvector1}) with the correct
normalization factor (\ref{norm}). This eigenstate
(\ref{eigenvector1}) is the first excited state (eigenstate of
neutron), not the vacuum in the whole Hilbert space $\mathcal{B}
\otimes \mathbb{C}^2$. Also, it is easy to see that the zero
eigenvalue of the Hamiltonian corresponds to the proton
state $\left(%
\begin{array}{c}
   |0 \rangle   \\
  0 \\
\end{array}%
\right)$.
Here $m>\mu$ for bound states since we want this model to describe
the attractive interaction of such a two state system with bosons.
However, it is not clear whether the proton is the state of lowest
energy. There may be states which contain many bosons that have a
lower energy. These questions will be answered in studying the
principal operator as we will do in the next section.

One can also generalize these ideas into the relativistic regime
with coupling constant renormalization, but this needs further
investigations so we are not going to elaborate on this and will
only consider the non-relativistic Lee model.

\section{A Lower Bound on the Ground State Energy}

After the renormalization of our model, we must look at the
spectrum of the problem because there are many theories in which
even after the renormalization there are still divergences that
makes the spectrum not bounded from below \cite{rajeevbound}. In this section we will restrict $E$  to the real axis. In
order to give the proof that the energy $E$ is bounded from below,
following the same idea as in \cite{rajeevbound}, we first split
the principal operator as
\begin{equation}
\Phi(E) = K(E) - U(E) \;,
\end{equation}
such that
\begin{equation}
K(E)= H_0 - E + \mu \;,
\end{equation}
and
\begin{eqnarray}
U(E) & = & - \lambda^2 \int_0^\infty d s \;
K_{s/2m}(a,a)\left[e^{-s(m-\mu)} -e^{-s(H_0+m-E)}\right]\cr & \ &
\hskip-1cm + \lambda^2 \int_0^\infty d s \int_{\mathcal{M}} d_g x
\, d_g y \; K_{s/2m}(x,a) \, K_{s/2m}(y,a) \, \phi^\dag(x)\,
e^{-s(H_0+2m-E)} \, \phi(y) \;. \end{eqnarray}
It follows immediately that $K(E) \geq n m - E + \mu $, so it is a
positive definite operator from our assumption $E< n m + \mu$. Due
to the positive definiteness of heat kernel $K_{s/2m}(a,a)>0$, the
difference of the two exponentials is a positive operator. As a
consequence of this, the first integral term in $U(E)$ is a
positive operator and we can claim that
\begin{equation} \nonumber
U(E) < U'(E) \;,
\end{equation}
where
\begin{equation} \nonumber
U'(E) = \lambda^2 \int_0^\infty d s \int_{\mathcal{M}} d_g x \,
d_g y \; K_{s/2m}(x,a) \, K_{s/2m}(y,a) \, \phi^\dag(x) \,
e^{-s(H_0+2m-E)} \, \phi(y) \;.
\end{equation}
This clearly forces
\begin{equation} \nonumber
\Phi(E) > K(E)
 - U'(E) \;,
\end{equation}
or rewriting it as
\begin{equation}
 \Phi(E) > K(E)^{1/2} \; \left(1-
 \tilde{U'}(E)\right) \; K(E)^{1/2}\;,
\end{equation}
where $\tilde{U'}(E) = K(E)^{-1/2}\; U'(E) \; K(E)^{-1/2}$ and
$K(E)$, $U'(E)$ are positive operators (so is $\tilde{U'}(E)$). We
will now show that by choosing $E$ sufficiently large enough it is
always possible to make the operator $\Phi(E)$ strictly positive,
hence it is invertible, and has no zeros beyond this particular
value of $E$. Therefore, if we impose
\begin{equation}
||\tilde{U'}(E)|| < 1 \;,
\end{equation}
then the principal operator $\Phi(E)$ becomes strictly positive.
In order to do some estimates, we will rewrite the interaction
term in terms of eigenfunctions $f_\sigma(x)$ and shift the
operator $\phi^{\dag}(x)$ to the leftmost side and the operator
$\phi(x)$ to the rightmost side
\begin{eqnarray}  \tilde{U'}(E) &=& \lambda^2\sum_{\sigma_1,\sigma_2}\phi^\dag
(\sigma_1) \, f_{\sigma_1}(a) \, [H_0+\mu+\sigma_1/2m+m-E]^{-1/2}
\cr & & \hskip-2cm \times \;
[H_0+\sigma_1/2m+\sigma_2/2m+2m-E]^{-1}
[H_0+\mu+\sigma_2/2m+m-E]^{-1/2} \, \phi(\sigma_2) \,
f_{\sigma_2}(a) \;, \nonumber
\end{eqnarray}
where $\phi(\sigma) = \int_{\mathcal{M}} d_g x \;
f^{*}_{\sigma}(x)\, \phi(x)$. In order to convert the product of
operators in the above formula into a  summation of them, we will
use the Feynman parametrization \cite{peskin}
\begin{eqnarray}
{1\over
A_1^{\alpha_1}A_2^{\alpha_2}A_3^{\alpha_3}}={\Gamma(\alpha_1+\alpha_2+\alpha_3)\over
\Gamma(\alpha_1) \Gamma(\alpha_2) \Gamma(\alpha_3) }\int_0^1
\Pi_i^3 du_i { \delta\left(\sum_i u_i-1 \right) \,
u_1^{\alpha_1-1}u_2^{\alpha_2-1}u_3^{\alpha_3-1}\over
[u_1A_1+u_2A_2+u_3A_3]^{\alpha_1+\alpha_2 +\alpha_3}} \;,
\end{eqnarray}
so that,
\begin{eqnarray}
\tilde{U'}(E)&=& \lambda^2 \sum_{\sigma_1, \sigma_2}
\phi^\dag(\sigma_1) \, f_{\sigma_1}(a)\, {\Gamma(1/2+1/2+1) \over
\Gamma(1/2)\Gamma(1/2)\Gamma(1)} \cr  &\ & \hspace{2.5cm} \times
\int_0^1 du_1 \, du_2 \, du_3 \; u_1^{{1 \over 2}-1} u_2^{{1 \over
2}-1} u_3^{1-1}
 \delta(u_1+u_2+u_3-1)\cr
 & \  & \hskip-1cm \times \; {1 \over
[H_0+m+\mu-E+(u_1+u_3){\sigma_1\over 2m}+(u_2+u_3){\sigma_2\over
2m}+
 u_3(m-\mu)]^2}\, \phi(\sigma_2) \, f_{\sigma_2}(a) \;, \nonumber
\end{eqnarray}
or one can rewrite it as
\begin{eqnarray} \nonumber
\tilde{U'}(E)&=& \lambda^2 \sum_{\sigma_1,\sigma_2}
\phi^\dag(\sigma_1)f_{\sigma_1}(a){\Gamma(2)\over
\Gamma(1/2)^2}\int_0^1 {du_1 \, du_2 \, du_3 \;
\delta(u_1+u_2+u_3-1) \over (u_1 \,u_2)^{1/2}}\cr &\ & \hskip-1cm
\times \int_0^\infty s \, d s \;
e^{-s(H_0+m+\mu-E+u_3(m-\mu)+(u_1+u_3)(\sigma_1/2m)+(u_2+u_3)
(\sigma_2/2m))} \, \phi(\sigma_2) \, f_{\sigma_2}(a) \;.
\end{eqnarray}
Let us express this equation in terms of the heat kernel:
\begin{eqnarray} \nonumber
\tilde{U'}(E)&=&  {\lambda^2 \, \Gamma(2)\over \Gamma(1/2)^2}
\int_0^\infty s \, d s \int_{\mathcal{M}} d_g x \, d_g y \;
\int_0^1 {du_1 \, du_2 \, du_3 \; \delta(u_1+u_2+u_3-1) \over (u_1
\,u_2)^{1/2} } \cr &\ & \hskip-1cm \times  \, \phi^\dag(x) \,
K_{s(u_1+u_3)/2m}(x,a) \, K_{s(u_2+u_3)/2m}(y,a) \,
e^{-s(H_0+\mu+m-E)} \, e^{-s u_3 (m-\mu)} \, \phi(y) \;.
\end{eqnarray}
It is easy to see that heat kernel has the following scaling
property in 3 dimensions:
\begin{equation}
K_{\alpha^2s}(x,y;g)=\alpha^{-3}K_s(x,y;\alpha^{-2}g) \;,
\end{equation}
where $g \rightarrow \alpha^{-2}g$ means that the metric $g_{ij}$
is scaled by a conformal factor $\alpha^{-2}$. Thus we have
\begin{eqnarray} \nonumber
\tilde{U'}(E)&=& {\lambda^2 \, \Gamma(2)\over \Gamma(1/2)^2}
\int_0^\infty s \, d s \int_{\mathcal{M}} d_{(u_1+u_3)^{-1}g}x \,
d_{(u_2+u_3)^{-1}g}y
 \cr &\ & \hspace{0.5cm}  \times \; \int_{0}^{1} {du_1 \, du_2 \, du_3 \; \delta(u_1+u_2+u_3-1) \over (u_1
\,u_2)^{1/2} } \, \phi^\dag(x) \,
K_{s/2m}\left(x,a;(u_1+u_3)^{-1}g \right) \cr &\ & \hskip-1cm
\times \;
 K_{s/2m}\left(y,a;(u_2+u_3)^{-1}g \right)\,
e^{-s(H_0+\mu+m-E)} \, e^{-su_3(m-\mu)} \, \phi(y) \;.
\end{eqnarray}
In addition, under the scaling of metric, the commutation
relations obey the following rule
\begin{equation}
[\phi_{\alpha^2g}(x),
\phi^\dag_{\alpha^2g}(y)]=\delta_{\alpha^2g}(x,y) \;,
\end{equation}
or
\begin{equation}
[\alpha^{3/2}\phi_{\alpha^2g}(x),
\alpha^{3/2}\phi^\dag_{\alpha^2g}(y)]= \delta_g(x,y) \;.
\end{equation}
This lead us to find the scaling property of the creation and
annihilation operators $
\phi_{\alpha^2g}(x)=\alpha^{-3/2}\phi_g(x)$. Using
$\phi_{(u_1+u_3)^{-1}g}(x)=(u_1+u_3)^{3/4}\phi_g(x)$ we define the
creation and annihilation operators with respect to the new metric
and obtain

\begin{eqnarray} \nonumber
\tilde{U'}(E)&=& {\lambda^2 \, \Gamma(2)\over \Gamma(1/2)^2}
\int_0^\infty s\, d s \int_{\mathcal{M}} d_{(u_1+u_3)^{-1}g}x \,
d_{(u_2+u_3)^{-1}g}y  \cr &\ & \hspace{5cm}  \times \; \int_0^1
{du_1 \, du_2 \, du_3 \; \delta(u_1+u_2+u_3-1) \over (u_1 \,
u_2)^{1/2} (u_1+u_3)^{3/4}(u_2+u_3)^{3/4}}\cr &\ & \hskip-1cm
\times
  \; K_{s/2m}\left(x,a;(u_1+u_3)^{-1}g \right) \;
K_{s/2m}\left(y,a;(u_2+u_3)^{-1}g \right)\cr &\ & \hspace{2cm}
\times \; \phi_{(u_1+u_3)^{-1}g}^\dag(x) \; e^{-s(H_0+\mu+m-E)} \;
e^{-su_3(m-\mu)} \phi_{(u_2+u_3)^{-1}g}(y) \;.
\end{eqnarray}
In order to give an upper bound estimate on the norm of the
operator $\tilde{U'}(E)$, we apply the Cauchy-Schwartz inequality
in the norm corresponding to the new metric and get
\begin{eqnarray} \nonumber
||\tilde{U'}(E)|| &<& n \; { \lambda^2 \, \Gamma(2)\over
\Gamma(1/2)^2} \int_0^\infty s \, d s \; e^{-s(nm+\mu-E)} \int_0^1
{du_1 \, du_2 \, du_3\; \delta(u_1+u_2+u_3-1)\over (u_1 \,
u_2)^{1/2}(u_1+u_3)^{3/4}(u_2+u_3)^{3/4}} \cr &\ & \hskip-2cm
\times
 \;\left[ \int_{\mathcal{M}} d_{(u_1+u_3)^{-1}g}x \; K^2_{s/2m}\left(x,a;(u_1+u_3)^{-1}g \right)
 \right]^{1/2} \cr &\ & \hspace{2.5cm} \times \left[\int_{\mathcal{M}}
 d_{(u_2+u_3)^{-1}g}y\;
K^2_{s/2m}\left(y,a;(u_2+u_3)^{-1}g \right) \right]^{1/2} \;.
\end{eqnarray}
Here we have replaced the term $H_0+m+\mu-E$ in the exponent by
its minimum value $nm+\mu-E$ and dropped the term
$e^{-su_3(m-\mu)} < 1$. By using the reproducing property of the
heat kernel, we get
\begin{eqnarray} \nonumber
||\tilde{U'}(E)||&<& n \; { \lambda^2 \, \Gamma(2)\over
\Gamma(1/2)^2} \int_0^\infty s\, d s \; e^{-s(nm+\mu-E)} \int_0^1
{du_1du_2du_3 \over (u_1 \,
u_2)^{1/2}(u_1+u_3)^{3/4}(u_2+u_3)^{3/4}} \cr &\ & \hskip-2.5cm
\times \delta(u_1+u_2+u_3-1)
 \left[ K_{s/m}(a,a;(u_1+u_3)^{-1}g) \right]^{1/2}\left[ K_{s/m}(a,a;(u_2+u_3)^{-1}g)
 \right]^{1/2} \;.
\end{eqnarray}
Scaling back to the original variables, we finally obtain
\begin{eqnarray}
||\tilde{U'}(E)||&<& n \; { \lambda^2 \, \Gamma(2)\over
\Gamma(1/2)^2} \int_0^\infty s \, d s \; e^{-s(nm+\mu-E)} \int_0^1
{du_1 \, du_2 \, du_3 \over (u_1 \, u_2)^{1/2}}
 \delta(u_1+u_2+u_3-1) \cr
&\ & \hspace{2cm} \times  \left[ K_{s(u_1+u_3)/m}(a,a;g)
\right]^{1/2} \left[ K_{s(u_2+u_3)/m}(a,a;g) \right]^{1/2}\;.
\end{eqnarray}
It is essential here to note the behavior of the heat kernel. In
most of the situations, the explicit form of the heat kernel is
unknown so one should look for the estimates or upper bounds on
it. Luckily, there are quite sharp upper bounds on the heat kernel
for various classes  of manifolds in the mathematical literature
\cite{grigoryan,liyau,mckean,wang}. For each class of manifolds,
there are
 different bounds so we will consider them separately.

We will first consider Cartan-Hadamard manifolds, which are
geodesically complete simply connected non-compact Riemannian
manifolds with non-positive sectional curvature $-K^2$ (for
example $\mathbb{R}^d$ and $\mathbb{H}^d$). On 3 dimensional
Cartan-Hadamard manifolds, we have the following upper bound on
the heat kernel (see chapter 7.4 in \cite{grigoryan} and
\cite{mckean})
\begin{equation}
K_{s/2m} (x,x) \leq {C \over \min \left\{{1 \over 2m},
(s/2m)^{3/2}\right\}} e^{-K^2 s /2 m} \;,
\end{equation}
or for simplicity of our calculations, one can use the following
less sharp estimate \cite{grigoryan}
\begin{equation} \label{cartanestim}
K_{s/2m} (x,x) \leq {C \over (s/2m)^{3/2}} \;,
\end{equation}
for all $x\in \mathcal{M}$ and $s>0$, and $C$ is a dimensionless
constant. This estimate (\ref{cartanestim}) can be extended also
to the minimal submanifolds of $\mathbb{R}^3$ \cite{grigoryan}.
Then the strict positivity of the principal operator after taking
the integrals leads to the following inequality
\begin{equation} \nonumber
||\tilde{U'}(E)||< n \; \lambda^2 \, m^{3/2} \, C \; { \pi^{3/2}
\Gamma(2) \Gamma(1/4)^2 \over \Gamma(1/2)^2 \Gamma(3/4)^2} \, (n m
+ \mu -E)^{-1/2} < 1 \;.
\end{equation}
This implies the inequality for $E$ or the opposite inequality for
the ground state energy written in ordinary units
\begin{equation}
E< n m c^2 + \mu c^2 - n^2 {\tilde{C}^2\lambda^4  m^3 \over
\hbar^6} \;, \ \quad \quad E_{gr} \geq n m c^2 + \mu c^2 - n^2
{\tilde{C}^2 \lambda^4 m^3 \over \hbar^6} \label{energycartan}\;,
\end{equation}
where %
\begin{equation} \nonumber
\tilde{C} = C \; { \pi^{3/2} \Gamma(2) \Gamma(1/4)^2 \over
\Gamma(1/2)^2 \Gamma(3/4)^2} \;.
\end{equation}
Secondly, as an explicit application, we consider the 3
dimensional hyperbolic space $\mathbb{H}^3$. The heat kernel in
$\mathbb{H}^3$ is explicitly known \cite{heatkernelh3} and it is
in natural units
\begin{equation} \label{heatkernelh3}
K_{s/2m}(x,y)= {1 \over R^3}{d(x,y)/R \over \sinh (d(x,y) /R)}
{e^{-{s \over 2 m R^2}-{m d^2(x,y) \over 2 s}}\over \left(4\pi {s
\over 2 m R^2}\right)^{3/2}} \;,
\end{equation}
where $R$ is the length scale and $d(x,y)$ is the geodesic
distance between two points on $\mathbb{H}^3$. Again we impose the
strict positivity condition on the principal operator and get
\begin{eqnarray} \nonumber
||\tilde{U'}(E)||&<& n \;  {\lambda^2 \, \Gamma(2)\over
\Gamma(1/2)^2} \int_0^1 {du_1 \, du_2 \, \over (u_1 \, u_2)^{1/2}
(1-u_1)^{3/4} (1-u_2)^{3/4}} \cr &\ & \hspace{2cm} \times \,
\int_{0}^{\infty}{ s \, d s \; e^{-s(nm+\mu-E)} e^{-s(1-u_1)/2m
R^2} e^{-s(1-u_2)/2m R^2} \over (4 \pi s /m)^{3/2}} < 1\;.
\end{eqnarray}
Using the fact that $e^{-s(1-u_{1(2)})/2mR^2} < 1$ and then taking
the integrals, one can immediately see that this implies
\begin{equation}
E_{gr} \geq n m c^2 + \mu c^2 - n^2 {\lambda^4 {D}^2 m^3 \over
\hbar^6} \label{energyh3}\;,
\end{equation}
where the constant $D$ is defined as
\begin{equation} \nonumber
D = {\Gamma(2) \Gamma(1/4)^2 \over  \Gamma(1/2)^2 \Gamma(3/4)^2
2^{3}} \;.
\end{equation}

Finally, we apply our method to the closed compact manifolds with
Ricci curvature bounded from below by $-\kappa$. The estimate for
this class of manifolds on the heat kernel \cite{liyau,wang} is
given by

\begin{equation} \label{heatkernelcompact}
K_{s/2m}(a,a) \leq {1 \over V(\mathcal{M})} + A (s/2m)^{-3/2} \;,
\end{equation}
where $A = A(V(\mathcal{M}), \kappa, d)$ is an explicitly
calculable constant which depends on the volume   $V(\mathcal{M})$ of the manifold,
 the lower bound $\kappa$ on the Ricci curvature 
and the diameter $d$ of the manifold. Using this estimate, it
follows that
\begin{eqnarray} \nonumber
||\tilde{U'}(E)||&<& n \;  {\lambda^2 \, \Gamma(2)\over
\Gamma(1/2)^2} \int_0^\infty s \, d s \; e^{-s(nm+\mu-E)} \int_0^1
{du_1 \, du_2 \, \over (u_1 \, u_2)^{1/2}} \cr &\ & \hskip-2.5cm
\times \left[ {1 \over V(\mathcal{M})^{1/2}} + A^{1/2}
(s(1-u_1)/m)^{-3/4}
 \right]   \left[ {1 \over V(\mathcal{M})^{1/2}} + A^{1/2} (s(1-u_2)/m)^{-3/4}
\right]\;.
\end{eqnarray}
Integrating with respect to $u_1$, $u_2$ and $s$, we have
\begin{eqnarray} \nonumber
||\tilde{U'}(E)||&<& n \; {\lambda^2 \, \Gamma(2)\over
\Gamma(1/2)^2} \bigg[{4 \over V(\mathcal{M})}{1 \over (n m + \mu
-E)^2} + {4 A^{1/2} m^{3/4} \pi^{1/2} \Gamma(1/4) \over V^{1/2}
\Gamma(3/4)} \cr &\ & \hskip-1cm \times \; {1 \over (n m + \mu
-E)^{5/4} } + {A m^{3/2} \pi \Gamma(1/4)^2 \over \Gamma(3/4)^2}{1
\over (n m + \mu -E)^{1/2} }
 \bigg]\;.
\end{eqnarray}
In order to get explicit estimates, let us put some further
assumption $nm + \mu - E > \mu$. Then, we find
\begin{eqnarray} \nonumber
||\tilde{U'}(E)||&<& n \; {\lambda^2 \, \Gamma(2)\over
\Gamma(1/2)^2} \bigg[{4 \over V(\mathcal{M}) \mu^{3/2}} + {4
A^{1/2} m^{3/4} \pi^{1/2} \Gamma(1/4) \over \mu^{3/4}
V(\mathcal{M})^{1/2} \Gamma(3/4)} + {A m^{3/2} \pi \Gamma(1/4)^2
\over \Gamma(3/4)^2}
 \bigg] \cr &\ & \times \; {1 \over (n m + \mu - E)^{1/2}} \;.
\end{eqnarray}
Now if we impose the strict positivity of the principal operator
we find similarly
\begin{equation}
E_{gr} \geq n m c^2 + \mu c^2 - n^2 \lambda^4 {F}^2
\label{energyh4}\;,
\end{equation}
where
\begin{equation} \nonumber
F = {\Gamma(2)\over \Gamma(1/2)^2}  \bigg[{4 \over V(\mathcal{M})
\mu^{3/2}} + {4 A^{1/2} m^{3/4} \pi^{1/2} \Gamma(1/4) \over
\mu^{3/4} V(\mathcal{M})^{1/2} \Gamma(3/4)} + {A m^{3/2} \pi
\Gamma(1/4)^2 \over \Gamma(3/4)^2}
 \bigg] \;.
\end{equation}
Therefore, lower bound on the ground state energies for different
classof manifolds (\ref{energycartan}), (\ref{energyh3}) and
(\ref{energyh4}) are of almost the same form up to a constant
factor so the form of the lower bound has a general character. It
is also worth pointing out that the form of the lower bound on the
ground state energy is same as in the case for the flat space
$\mathbb{R}^3$ \cite{rajeevbound,ali}. From the general form of
the lower bounds, we conclude that for each sector with a fixed
number of bosons, there exists a ground state. However, the ground
state energy bound that we have found diverges quadratically as
the number of bosons increases. In other words, these estimates
with our present analysis is not good enough to prove the
existence of the thermodynamic limit. To attack this problem we
will study the thermodynamic limit of the model by mean field
approximation.

\section{Mean Field Approximation of the Model}

In the limit of large number of bosons $n\rightarrow \infty$, one
expects that all the bosons have the same wave function $u(x)$ and
mean field approximation is valid, as in the case of flat spaces.
Therefore, one can introduce the following mean field ansatz for
$n$ particle state on a Riemannian manifold
\begin{equation}
|u \rangle = {1 \over \sqrt{n!}} \int_{\mathcal{M}} d_g x_1 \, d_g
x_2 \, \cdots d_g x_n \; u(x_1)\,u(x_2)\cdots u(x_n)\,
\phi^{\dag}(x_1)\, \phi^{\dag}(x_2) \cdots  \phi^{\dag}(x_n) |0
\rangle \;,
\end{equation}
with the normalization
\begin{equation} \label{normalization}
||u(x)||^2 = \int_{\mathcal{M}} |u(x)|^2 \; d_g x = 1 \;,
\end{equation}
where $|0\rangle$ denotes the vacuum state. In the mean field
approximation the operators are usually approximately replaced by
their expectation values in this state ($\langle f(u) \rangle
\approx f(\langle u \rangle)$) so the expectation value of the
principal operator by applying the mean field ansatz becomes
\begin{eqnarray}
\phi(E,u)&=&n h_0(u)-E+\mu+\lambda^2\int_0^\infty d s \;
K_{s/2m}(a,a)\,[e^{-s(m-\mu)}-e^{-s(nh_0(u)+m-E)}]\cr &\ &
\hskip-2cm - \; n \lambda^2\int_0^\infty d s \int_{\mathcal{M}}
d_g x \, d_g y \; K_{s/2m}(x,a)\, K_{s/2m}(y,a)\, u^*(x)\,
e^{-s(nh_0(u)+2m-E)} \, u(y) \;, \label{principalfunc}
\end{eqnarray}
called principal function and we have defined
\begin{equation}
h_0 (u)= \int_{\mathcal{M}} d_g x \; \left( { |\nabla_g u(x)|^2
\over 2 m} + m |u(x)|^2 \right) \;.
\end{equation}
However, the exact value of the expectation value of the principal
operator is given in terms of cummulant expansion theorem
\cite{Ma}. Therefore, in order to write the above formula
(\ref{principalfunc}), we have to further assume that the
corrections coming from the higher order cummulants are negligibly
small and indeed we will see that this assumption is justified for
the particular solution we will find.

Now, we must solve the equation $\phi(E,u) =0$ (due to
(\ref{spectrum})) in order to find the spectrum of the problem and
solve $E$ as a function of $u(x)$, which gives the smallest
possible value of $E$ with the constraint (\ref{normalization}).
Hence, one can try to write $E$ as a functional of $u(x)$ from the
equation $\phi(E,u) =0$ and apply the variational methods to
minimize $E$.

However, there is no simple way to solve this variational problem
. So, we follow a different method essentially the one suggested
in \cite{rajeevbound}. For this purpose, let us introduce a new
variable $\chi$ and new wave function $v(x)$
\begin{eqnarray} \chi & = & n h_0(u)-E \cr
v(x)&=&[2m(2m+\chi)]^{-3/4}u(x) \;,
\end{eqnarray}
such that the new wave functions $v(x)$ are normalized with
respect to the new metric $\tilde g_{ij}=[2m(2m+\chi)] g_{ij}$
\begin{equation}
\int_{\mathcal{M}} d_{\tilde g} x \; |v(x)|^2= \int_{\mathcal{M}}
d_{g} x \; |u(x)|^2 = 1 \;.
\end{equation}
We also define a new dimensionless parameter $s' = (2m+\chi)s$,
and using the scaling property of heat kernel we find
\begin{eqnarray}
\chi+\mu+\lambda^2\int_0^\infty d s \;
K_{s/2m}(a,a)\left[e^{-s(m-\mu)}-e^{-s(\chi+m)}\right] & & \cr  &
\ & \hskip-8cm = n\lambda^2(2m)^{3/2} (2m+\chi)^{1/2}
\int_0^\infty d s' \; \left| \int_{\mathcal{M}} d_{\tilde g}x \;
K_{s'}(x,a;\tilde g) \, v(x)\right|^2 e^{-s'} \;.
\label{meanfieldeq}
\end{eqnarray}
One can prove now that the left hand side as a function of the
variable $\chi$ is an increasing function and it is obvious that
the righthand side is positive. Therefore, the left hand side is
minimum when $\chi=-\mu$, which is attained when the right hand
side becomes zero (so $\chi \geq -\mu$). Let us denote the inverse
function of the left hand side as $f_1(nU)$, that is,
\begin{equation}
\chi = f_1(nU) \;,
\end{equation}
and express $\chi$ in terms of the energy $E$ and the function
$v(x)$,
\begin{equation}
   \chi=n [\chi + 2 m]K[v]+ n m - E, \ \ \ {\rm or}
\end{equation}
\begin{equation} \label{energy f1}
E=nm+ 2m nK[v]+(n K[v]-1) f_1(nU) \;,
\end{equation}
where $K[v]=\int d_{\tilde g} x |\nabla_{\tilde{g}} v(x)|^2$,
which is considered to be the parameter of the model because it is
the variable we can control and
 we may use many trial functions $v(x)$ and
they can be scaled to any desired value. If we assume that $n
K[v]>1$, then the energy $E$ (\ref{energy f1}) is minimized when
$f_1(n U)$ is minimized which happens for $U[v]=0$. Since $\chi =
f_1(n U) \geq - \mu$, we have
\begin{equation}
E \geq n m + \mu \hspace{1cm} \mathrm{if} \hspace{1cm}  n K[v]>1
\;.
\end{equation}
On the other hand, if $K[v]$ is small enough, i.e., $n K[v]<1$, we
also see that the minimum of the energy is attained with the
reversed sign of the last term. In that case, we should find an
upper bound for $f_1(nU)$ which is expressed in terms of the
kinetic energy functional $K[v]$. In order to discuss the case $n
K[v]<1$ properly, we will separate our calculations for compact
and non-compact manifolds.

Let us first consider the case for compact manifolds. In order to
achieve our aim, we will go back to the original variable $u(x)$
and the parameter $s$ in the equation (\ref{meanfieldeq})
\begin{eqnarray}
\chi+\mu+\lambda^2\int_0^\infty d s \;
K_{s/2m}(a,a)\left[e^{-s(m-\mu)}-e^{-s(\chi+m)}\right] & & \cr  &
\ & \hskip-8cm = n\lambda^2 \int_0^\infty d s \; \left|
\int_{\mathcal{M}} d_{g} x \; K_{s/2m}(x,a ; g) \, u(x)\right|^2
e^{-s(\chi + 2 m)} = n U[u] \label{meanfieldeqor}\;.
\end{eqnarray}
If we assume that  the mean field approximation gives us a
reliable equality, we can  find the solution for $\chi$ given all
the other parameters. Note that the right hand side of
(\ref{meanfieldeqor}) is a decreasing function of $\chi$ whereas
the left hand side is an increasing one. Hence there is always a
unique solution which defines the inverse function for a given
$u(x)$. It is quite easy to see by a graphical construction that
if we replace the left hand side by a smaller function of $\chi$,
and the right hand side by a larger function of $\chi$, then we
get an upper bound for the inverse function. We  note first that
\begin{eqnarray} \nonumber
\int_{\mathcal{M}} d_{g} x \; K_{s/2m}(x,a; g) \, u(x) =
\sum_{\sigma} e^{-{s \sigma \over 2m}} f_\sigma(a) \, u(\sigma)
\;,
\end{eqnarray}
where $u(\sigma)= \int_{\mathcal{M}} d_g x \; f_{\sigma}^{*}(x)\,
u(x)$. If we define the following functions
\begin{equation}
 f'_{\sigma}  (a) =
\begin{cases}
\begin{split}
{2 m f_{\sigma}  (a) \over \sqrt{\sigma}}
\end{split}
& \quad \textrm{if $\sigma \neq 0 $} \\ \\
\begin{split}
f_{0}(a)
\end{split}
& \quad \textrm{if $\sigma =0 $},
\end{cases}
\end{equation}
and
\begin{equation}
 u'(\sigma) =
\begin{cases}
\begin{split}
{\sqrt{\sigma} \, u(\sigma) \over 2m}
\end{split}
& \quad \textrm{if $\sigma \neq 0 $} \\ \\
\begin{split}
u(0)
\end{split}
& \quad \textrm{if $\sigma =0 $},
\end{cases}
\end{equation}
we can write
\begin{eqnarray} \nonumber
&\ & \left| \int_{\mathcal{M}} d_{g} x \; K_{s/2m}(x,a; g) \, u(x)
\right|^2  = \left| \sum_{\sigma} e^{- {s \sigma \over 2m} }
f'_{\sigma}(a) u'(\sigma)\right|^2  \cr &\ & \hskip-1cm \leq
\sum_{\sigma} e^{- {s \sigma \over m}} |f'_{\sigma}(a)|^2
\sum_{\sigma} |u'(\sigma)|^2 \leq \left( |f_0 (a)|^2 + 2 m
\sum_{\sigma}^{'} {e^{-{s \sigma \over m}}\over \sigma/2m}
|f_{\sigma}(a)|^2 \right) \left( |u(0)|^2 + {K[u]\over 2m} \right)
\;,
\end{eqnarray}
where $\sum_{\sigma}^{'}$ is the sum which excludes the zero mode
($\sigma=0$), $K[u]= {1 \over 2m} \int_{\mathcal{M}} d_g x
|\nabla_g u(x)|^2 $, and we have used Cauchy-Schwartz inequality.
Using $|f_0 (a)|^2 = 1/ V(\mathcal{M}) $ and $|u(0)|^2 \leq 1$, we
find
\begin{equation}
U[u] \leq   \left(1 + {K[u]\over 2m}\right) \; \Omega \;,
\label{U}
\end{equation}
where
\begin{eqnarray}
&\ & \Omega =  {\lambda^2 \over \chi+2m} \bigg[ {1 \over
V(\mathcal{M})} + 2 m \int_0^\infty d s \; \left(1-
e^{-s(\chi+2m)/2 }\right)\cr &\ & \hspace{5cm} \times\; \left(
K_{s/2m}(a,a;g) - \lim_{s \rightarrow \infty} K_{s/2m}(a,a; g)
\right) \bigg] \;.
\end{eqnarray}
Here the sum which excludes the zero mode corresponds to
substraction of  the large $s$ behavior of heat kernel. Using
$K[u]=(2m+\chi) K[v]$, the inequality becomes
\begin{eqnarray} \nonumber
&\ & \chi + \mu \leq {n \lambda^2 \over \chi+2m} \bigg [{1 \over
V(\mathcal{M})} +2 m \int_0^\infty d s \; \bigg(1-
e^{-s(\chi+2m)/2}\bigg) \cr &\ & \hspace{2.5cm} \times \; \left(
K_{s/2m}(a,a;g)  - \lim_{s \rightarrow \infty} K_{s/2m}(a,a; g)
\right) \bigg ] \left[ 1 +{(\chi + 2 m) \over 2 m} K[v]\right]\;.
\label{compactineq}
\end{eqnarray}
Using the upper bound estimate of the heat kernel for closed
compact manifolds with Ricci curvature bounded from below by
$-\kappa$ and taking the integral with respect to $s$, we find
that
\begin{eqnarray}
\chi+\mu&\leq & {n\lambda^2 \over \chi + 2 m} \left[ 1 + \left(
{\chi + 2 m \over 2 m}\right) K[v]\right] \Big[{1\over
V(\mathcal{M})} + \sqrt{2 \pi} (2m)^{5/2} A (\chi + 2 m)^{1/2}
\Big]\;,
\end{eqnarray}
If we now introduce the variables $z=\chi+2m$ and $A' = \sqrt{2
\pi} (2m)^{5/2} A $ for simplicity of notation, we find the
following inequality
\begin{equation}
z-(2m-\mu)\leq  { \lambda^2 \over z}\left(n +  {z \over 2 m}
\right)\left(A' z^{1/2}+ {1\over V(\mathcal{M})} \right) \;.
\end{equation}
We look for a systematic expansion of $z$ in $n$ by allowing a
fractional power. In this case we see that if we substitute the
following asymptotic expansion as $n\rightarrow \infty$, $ z \sim
B_1n^{\nu_1} + B_2 n^{\nu_2}+B_3 n^{\nu_3} + \cdots $, where the
consecutive powers $\nu_1,\nu_2,\ldots$ decrease, we find that
this asymptotic expansion is an upper bound for this inequality as
long as the coefficients are to be chosen as $B_1 = (A'
 \lambda^2 )^{2/3}$, $B_2 = {1 \over V(\mathcal{M})}
 \left({\lambda \over A'}\right)^{2/3}
 + {1 \over 2m} (A' \lambda^2)^{4/3}$
 and $B_3 = 2m - \mu + {\lambda^2 \over 2 m V(\mathcal{M})}$, where $\nu_1 = 2/3$, $\nu_2 = 1/3$, and $\nu_3 = 0$.
 Hence we get an upper bound on the inverse function $f_1(nU)$
as an asymptotic series in powers of $n$, in the spirit of mean
field approximation. This upper bound can be put back into the
energy equation and then we find the following lower bound on the
ground state energy,
\begin{equation}
E_{gr}\geq nm+ 2m n K[v]-\left(1-n K[v] \right)\left(
B_1n^{2/3}+B_2n^{1/3}+B_3-2m+\cdots \right)
\end{equation}
We note now that the behavior of $K[u]$ for large $n$ is found
from the scaling law, $K[u]=(2m+\chi)K[v]\sim n^{-1/3}$. This in
turn justifies our use of the mean field approximation since
higher order derivatives are then negligible, and the solution is
approaching to an essentially constant function on the manifold as
$n$ gets larger. This proves that for a compact manifold the
energy is actually bounded form below by a much milder behavior
and there is a nice thermodynamic limit since the energy per
particle $E/n$ will approach to the mass (rest mass energy) as
$n\rightarrow \infty$. The same conclusion can also be drawn for
non-compact manifolds as we will see.

Now we will consider the mean field approximation of the model for
non-compact manifolds. We again assume that the eigenfunction
expansion in compact manifolds can be generalized to the
non-compact manifolds. Then, one can use the above method for the
non-compact manifolds as well. However, we shall try to find the
ground state energy in the mean field approximation in an another
way. Therefore, we first go back to the equation:
\begin{eqnarray} \nonumber
{1\over (\chi + 2 m)^{1/2}}\left\{ \chi+\mu+\lambda^2\int_0^\infty
d s \; K_{s/2m}(a,a)
\left[e^{-s(m-\mu)}-e^{-s(\chi+m)}\right]\right\} & & \cr &\ &
\hskip-10cm =n\lambda^2(2m)^{3/2} \int_0^\infty d s' \left|
\int_{\mathcal{M}} d_{\tilde g}x \; K_{s'}(x,a;\tilde g)\,
v(x)\right|^2 e^{-s'} \equiv n \; U[v] \;,
\end{eqnarray}
and using the generalized eigenfunction expansions, we find
\begin{equation}
U[v] \leq K[v] \, \Omega \;,
\end{equation}
where
\begin{equation}
\Omega =  \lambda^2 (2m)^{3/2}  \int_0^\infty d s' \; \left(1-
e^{-s'/2}\right) K_{s'}(a,a;\tilde g) \;,
\end{equation}
The explicit form of the inverse function $f_1(n U)$ is too
difficult to find so one can estimate it. In order to do this, let
us first notice that
\begin{equation} \nonumber
 f_2^{-1} (\chi) < f_1^{-1} (\chi) \ \ \ \ {\rm then}\ \ \ f_1 (n U) < f_2 (n U)
.\end{equation}
For this purpose, we  use the simplest possible function as
$f_2^{-1} (\chi)$
\begin{equation} \nonumber
 f_2^{-1} (\chi) = {\chi+\mu \over (\chi + 2 m)^{1/2}} \;.
 \end{equation}
Then we can replace $f_1 (n U)$ with something bigger, and its
argument with something even bigger. Moreover,  $f_2 (u)$ is
dominated by a simpler function
\begin{equation} \nonumber
 f_2(u)< u^2+2m-2\mu,
\end{equation}
which  is a very crude bound, but easy to work with. Using the
upper bound for $U[v]$,  we get
\begin{equation}
 \chi = f_1 (n U) < f_2 (u) < n^2 U[v]^2+ 2m - 2 \mu < n^2 (K[v])^2 \Omega^2
 +2m-2\mu \;.
\end{equation}
Hence,
\begin{equation}
E \geq n m + 2 m n K[v] - \left(1-n K[v]\right)\left( n^2 K[v]^2
\, \Omega^2 + 2 m- 2 \mu  \right) \;,
\end{equation}
where the polynomial $(4m-2\mu) y - \Omega^2 y^2 +  \Omega^2 y^3$
{\it never becomes negative} within the range $0 \leq y = n K[v]
\leq 1$ if
\begin{equation} \label{omegacond}
\Omega^2< 16m-8\mu \;.
\end{equation}
In this case, the minimum is achieved when $y=0$. This means that
the functional $nK[v] \to 0$ by a proper family of functions. This
is why it is consistent to ignore the higher order cummulants if
we choose an arbitrarily slowly varying family for the function
$v(x)$. Therefore one can conclude that the ground state energy
for non-compact manifolds when $n K[v]<1$
\begin{equation}
E_{gr}\geq n m - 2(m-\mu)\;.
\end{equation}
The condition on $\Omega$ may imply some restrictions on the
coupling constant $\lambda$. If we go back to
the definition of $\Omega$ and scaling back again to the usual
geometric variables, we find
\begin{equation}
\Omega = \lambda^2 (\chi + 2 m)^{-1/2} \int_0^\infty d s \;
\left(1- e^{-s(\chi + 2 m)/2} \right) K_{s/2m}(a,a; g)  \;
.\end{equation}
Since there is a nice sharp upper bound on the heat kernel for
Cartan-Hadamard manifolds and minimal submanifolds of
$\mathbb{R}^3$ (\ref{cartanestim}), we can find an upper bound on
$\Omega$ and the restriction (\ref{omegacond}) gives the following
upper bound on the coupling constant.
\begin{equation}
\lambda < { (16 m - 8 \mu)^{1/4} \over (2\pi)^{1/4} C^{1/2}
(2m)^{3/4}} \;.
\end{equation}
Now, let us calculate explicitly the function $\Omega$ for the
hyperbolic manifold $\mathbb{H}^3$, which belongs to the class of
Cartan-Hadamard manifolds. Using the result (\ref{heatkernelh3}),
we obtain
\begin{equation}
\Omega = \lambda^2 \left(\chi + 2 m \right)^{-1/2} \left(4\pi / 2
m \right)^{-3/2} 2 \sqrt{\pi} \left\{\sqrt{{\chi + 2 m \over 2}+{1
\over 2 m R^2}} - \sqrt{{1 \over 2 m R^2}} \right\} \;.
\end{equation}
Then, one can easily find the upper bound on the coupling constant
from the restriction on $\Omega$
\begin{equation}
\lambda < 2 \sqrt{4 \pi (2m-\mu)} (2m)^{-3/4} \left\{ \sqrt{ {
2m-\mu \over 2 } + { 1 \over 2m R^2}} - \sqrt{{1 \over 2 m R^2}}
\right\}^{-1/2} \;.
\end{equation}
Therefore, similar to the case for compact manifolds, we have
shown that the leading behavior of the system varies linearly with
the number of bosons $n$, which leads to a nice thermodynamic
limit on non-compact manifolds.

\section{Conclusion}

In this paper, we considered the non-relativistic Lee model on
various class of Riemannian manifolds  inspired  from
the work in \cite{rajeevbound}. This method allows us to
renormalize the model non-perturbatively. It has been also shown
that the heat kernel plays a key role in the renormalization
procedure and help us to find a lower bound on the ground state
energy due to the sharp bound estimates on it for several class of
manifolds.  Finally, we studied the mean field approximation and
showed that there exist a nice thermodynamic limit of the model
for compact and non-compact manifolds.

\section{Acknowledgments}

The authors gratefully acknowledge the discussions of S. G.
Rajeev. O. T. T. 's research is partially supported by the Turkish
Academy of Sciences, in the framework of the Young Scientist Award
Program (OTT-TUBA-GEBIP/2002-1-18).

\end{document}